\documentclass[aps,twocolumn,superscriptaddress]{revtex4}
\usepackage[dvipdf]{graphicx}

\newcommand{\bra}[1] {\left\langle #1 \right|}
\newcommand{\ket}[1] {\left| #1 \right\rangle}
\newcommand{\braket}[2] {\left\langle #1 | #2 \right\rangle}

\begin{document}

\title{Attempt to find the hidden symmetry in the asymmetric quantum Rabi model}

\author{S. Ashhab}
\affiliation{Qatar Environment and Energy Research Institute, Hamad Bin Khalifa University, Qatar Foundation, Doha, Qatar}

\begin{abstract}
It has been observed that the asymmetric quantum Rabi model (QRM), which does not possess any obvious symmetry, exhibits energy level crossings, which are often associated with symmetries. This observation suggests that there is in fact a symmetry in the asymmetric QRM, even though simple inspection of the model and its Hamiltonian does not reveal the nature of this symmetry. Here we present the results of numerical calculations on the energy eigenstates of the asymmetric QRM in an attempt to elucidate the nature of the symmetry. In particular, we note that the distribution of states in the Hilbert space among different symmetry classes is normally independent of system parameters and test whether this property holds for the asymmetric QRM. We find that it does not, which both helps explain why the symmetry is hidden and adds more intrigue as to what its nature might be.
\end{abstract}

\maketitle

\section{Introduction}
\label{Sec:Introduction}

Symmetry is a fundamental concept in physics \cite{Gross}. It was used by the ancient Greeks to explain the balance in states of equilibrium in nature. In the middle ages, scientists for centuries assumed that the paths of celestial objects must be circular, because circular paths are the most symmetric ones possible, and this assumption might have delayed the discovery of the correct model of planetary motion. In the second half of the twentieth century, the violation of P and CP symmetries played an important role in the development of the standard model of particle physics and has been used to explain the matter-antimatter asymmetry in the universe.

Symmetries are closely related to conservation laws, as formulated in Noether's theorem \cite{Noether}. Identifying a symmetry can hence be a powerful tool that allows us to infer certain properties of a physical system even without knowing any of its quantitative parameters. For example, the laws of momentum and energy conservation can be inferred from the fact that nature possesses space and time translation symmetries. Another example of an interesting phenomenon resulting from symmetry is the phenomenon that has been predicted for Bose gases that the conservation law associated with spin rotation symmetry can prevent a cooled gas from reaching its ground state, which is a single Bose-Einstein condensate, leading to a situation in which the gas can only reach a double-condensate state in the spin-constrained ground state \cite{Kuklov,Ashhab2003}. In a somewhat opposite sense, the observation of spin-orbit interactions in a semiconductor material can be used to infer a lack of inversion (or mirror) symmetry in the material, because inversion-symmetric systems would not exhibit this type of spin-orbit coupling.

In the study of quantum systems, it is known that energy levels generally exhibit the phenomenon of avoided level crossings. Exceptions to this rule occur when there is a symmetry in the system under which there is no physical mechanism that would change the state of the system from one symmetry class to another symmetry class. In this case, quantum states that belong to different symmetry classes can exhibit energy level crossings, i.e.~situations in which two or more quantum states have exactly the same energy \cite{BooksOnSymmetryAndCrossings}.

An interesting physical model in relation to symmetry is the asymmetric quantum Rabi model (QRM) \cite{Irish,Niemczyk,Forn2010,Ashhab2010,Braak2011,Zhong,Larson,Tomka,Li,Forn2016,Batchelor,Yoshihara,Rossatto,Wakayama,
Kimoto,Semple,Guan,Mao,Wang,Braak2019}, which we shall describe in more detail below. This model does not seem to possess any symmetry. Nevertheless, it exhibits the phenomenon of energy level crossings at certain parameter values. This fact was conjectured in Ref.~\cite{Braak2011}, proved at different levels of generality in Refs.~\cite{Wakayama,Kimoto} and supported by the results of numerical studies \cite{Li,Kimoto}. This situation raises the question of what symmetry might exist in the system and hence explain the observed energy level crossings. Because its nature is unknown, the hypothesized symmetry has been referred to as a hidden symmetry.

Here we make an attempt to identify the hidden symmetry in the asymmetric QRM by numerically evaluating the energy eigenstates of the system at different parameter values and looking for a partitioning of the Hilbert space that is independent of the exact parameter values. We find that, unlike the symmetric QRM, no parameter-independent partitioning of the Hilbert space exists for the asymmetric QRM. These results can guide future investigations into the nature of the hidden symmetry in this model.

The remainder of this paper is organized as follows: In Sec.~\ref{Sec:RabiModel} we introduce the QRM, including its symmetric version. In Sec.~\ref{Sec:SearchingForPartition} we present our results on the energy eigenstates and draw our main conclusion about the lack of parameter-independent symmetry operators. In Sec.~\ref{Sec:Discussion} we discuss additional aspects of our results and, more generally, of the symmetry in the asymmetric QRM. In Sec.~\ref{Sec:NonIntegerEpsilonValues} we present results on the search for additional cases of hidden symmetry in the same model. We conclude with some final remarks in Sec.~\ref{Sec:Conclusion}.

\section{Quantum Rabi model and its symmetry}
\label{Sec:RabiModel}

The QRM describes a system composed of a qubit coupled to a harmonic oscillator with the Hamiltonian
\begin{equation}
\hat{H} = \frac{\Delta}{2} \hat{\sigma}_z + \frac{\epsilon}{2} \hat{\sigma}_x + \omega \left( \hat{a}^\dagger \hat{a} + \frac{1}{2} \right) + g \hat{\sigma}_x (\hat{a}+\hat{a}^\dagger),
\end{equation}
where $\Delta$ is the qubit gap, $\epsilon$ is the qubit bias parameter, $\omega$ is the oscillator's characteristic frequency (meaning that we set $\hbar=1$), $g$ is the qubit-oscillator coupling strength, the operators $\hat{\sigma}_{\alpha}$ (with $\alpha=x,y,z$) are the qubit's Pauli operators, and $\hat{a}$ and $\hat{a}^{\dagger}$ are, respectively, the annihilation and creation operators of the harmonic oscillator.

When $\epsilon=0$ the Hamiltonian has a symmetry described by the parity operator
\begin{equation}
\hat{\Pi} = \exp\left[ i\pi \left( \frac{1+\hat{\sigma}_z}{2} + \hat{a}^{\dagger}\hat{a} \right)\right].
\label{Eq:ParityOperator}
\end{equation}
The physical effect of this operator is to effect a $\pi$ rotation about the z axis in the state of the qubit and at the same time effect a $\pi$ rotation about the origin in the position-momentum phase space of the harmonic oscillator. In other words, it flips the state of the qubit in the $\sigma_x$ basis and at the same time creates the mirror image of the oscillator's wave function about the origin (i.e.~about the point $x=0$ in the real-space representation of the harmonic oscillator's wave function). If $\hat{\Pi}$ is applied to any eigenstate of the Hamiltonian, it produces the same state multiplied by the factor $\pm 1$. This property of the energy eigenstates can be seen straightforwardly in a standard representation of the corresponding wave functions: all energy eigenstate wave functions look either perfectly symmetric or antisymmetric, and the parity value of the state indicates whether the state belongs to the space of symmetric or antisymmetric states. This situation represents a $Z_2$ symmetry of the Hamiltonian.

\begin{figure}[h]
\includegraphics[width=7.0cm]{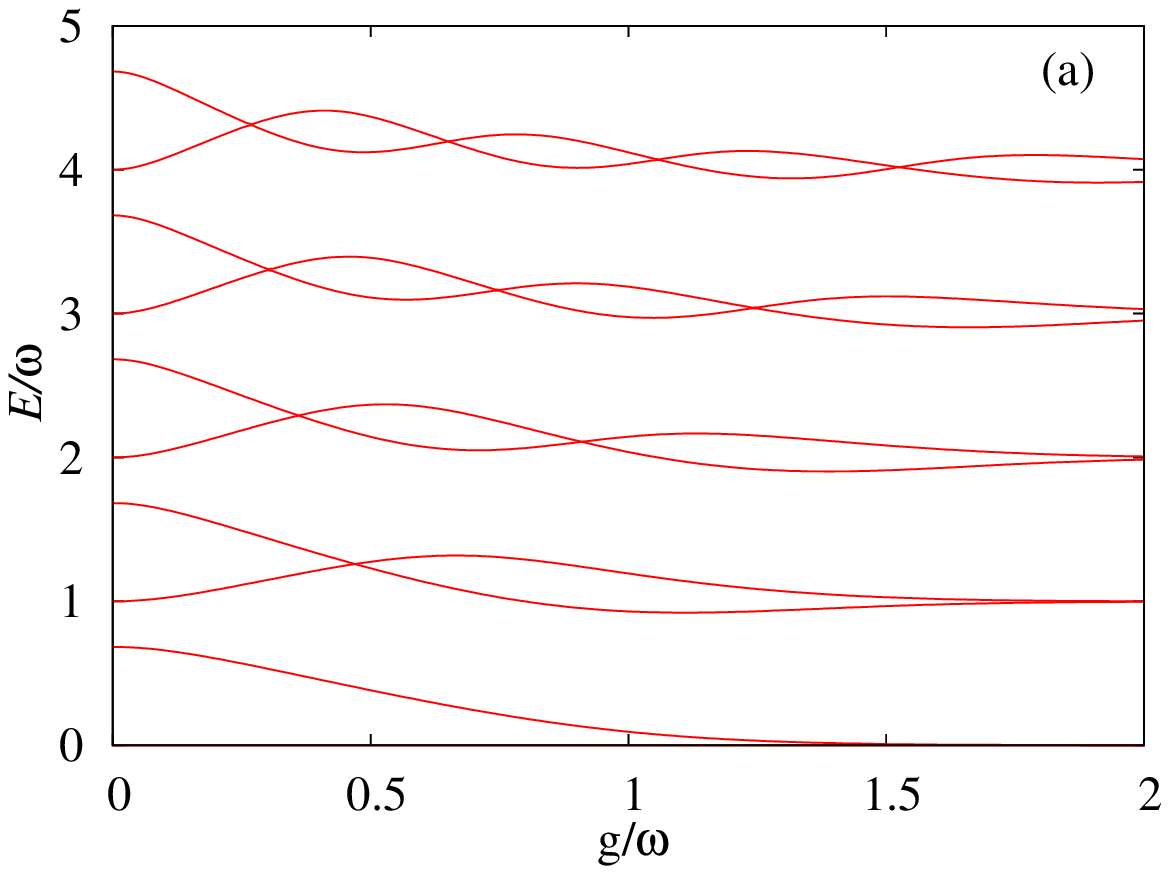}
\includegraphics[width=7.0cm]{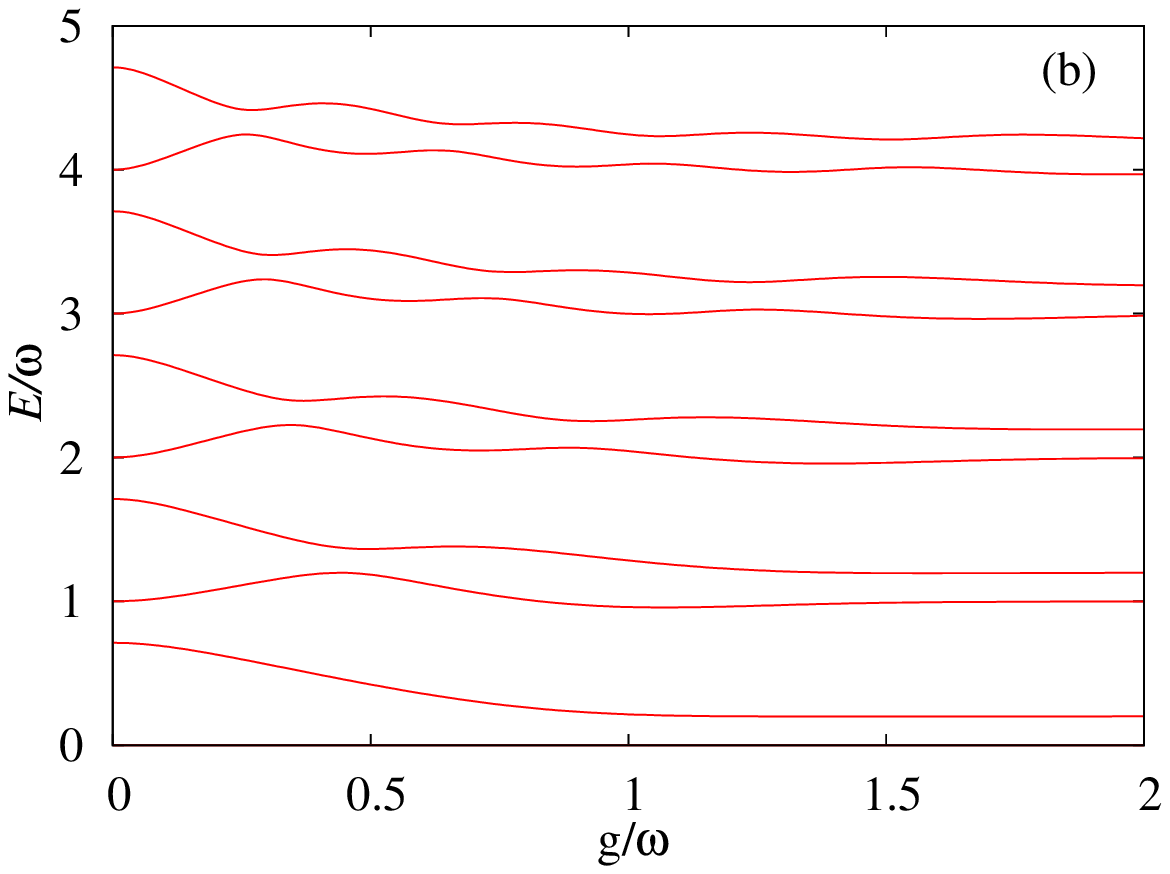}
\includegraphics[width=7.0cm]{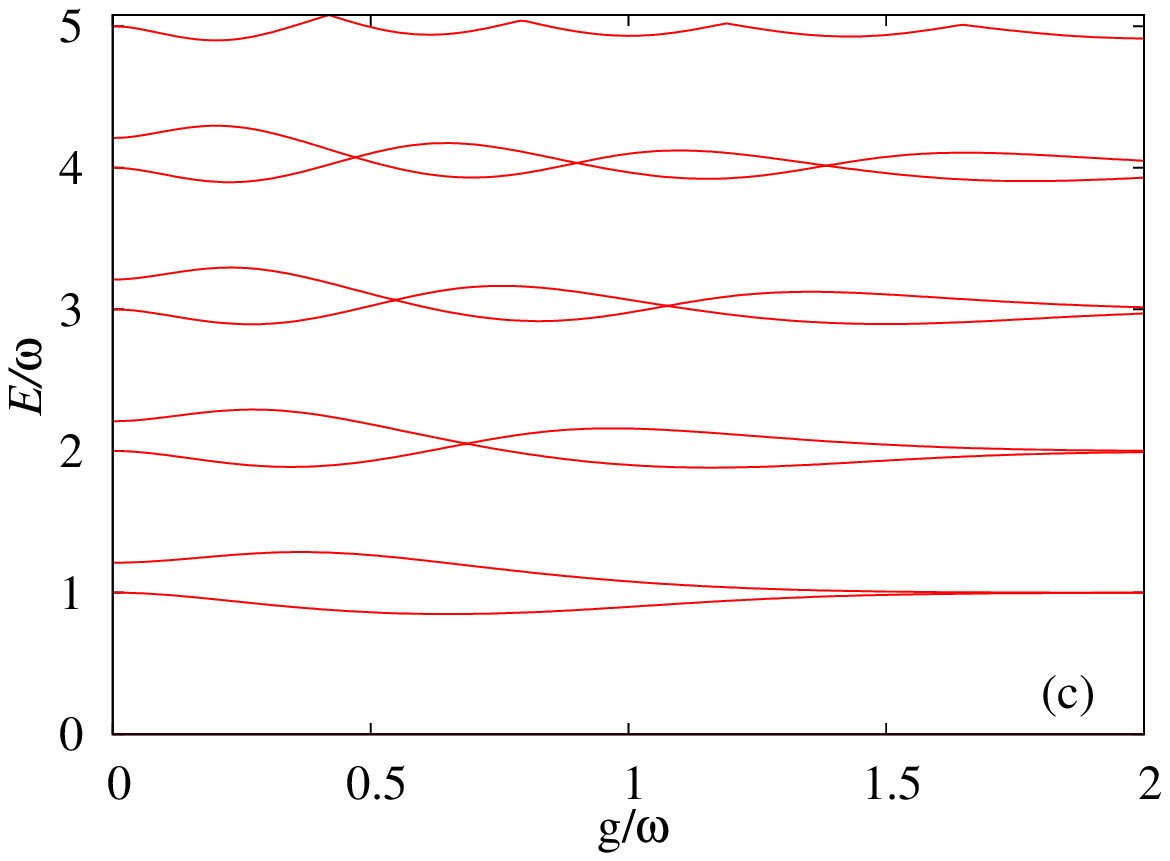}
\includegraphics[width=7.0cm]{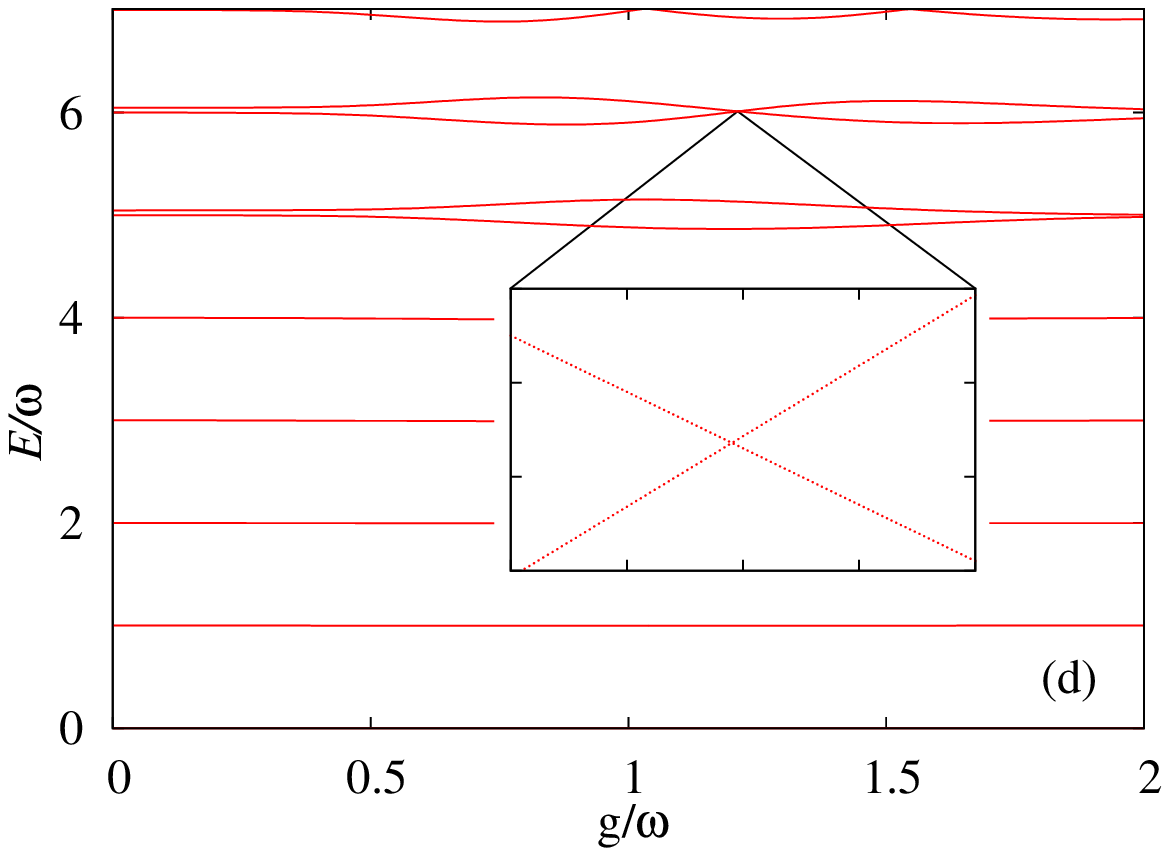}
\caption{Lowest ten energy levels (measured in units of $\omega$) as functions of coupling strength $g/\omega$ for different bias points: $\epsilon/\omega=0$ (a), 0.2 (b), 1 (c) and 5 (d). In all panels we measure the energy relative to the ground state energy, which therefore always coincides with the x axis. The inset in Panel (d) shows a magnified view of an energy level crossing region: the x axis range in the inset is $x_{\rm min}=1.21279135$ to $x_{\rm max}=x_{\rm min}+10^{-8}$, and the y axis range is $y_{\rm min}=6.011263858$ to $y_{\rm max}=y_{\rm min}+6 \times 10^{-9}$). Each dot in the inset is obtained from a separate calculation, i.e.~the dots are not simply a dotted line. The inset hence shows that the magnified structure is indeed a crossing and not an avoided crossing with a small gap (because with the parameter values used here we would expect any finite gap to be well above the $10^{-9}$ scale). In all Panels we set $\Delta/\omega=\pi^{-1/3}$, which is chosen because it is an irrational number that is somewhat close to 1.}
\label{Fig:EnergyLevels}
\end{figure}

A result of the symmetry in the Hamiltonian is that one can obtain energy level crossings when one or more of the system parameters are varied. For example, if $g$ is varied from zero to $\infty$ while $\Delta$ and $\omega$ are kept fixed, the energy levels exhibit many crossings, as can be seen in Fig.~\ref{Fig:EnergyLevels}(a).

In the asymmetric QRM, i.e.~when $\epsilon\neq 0$, the parity symmetry explained above is lost. In fact, there is no obvious symmetry remaining in the model. This fact can be seen by numerically evaluating the energy eigenstates and inspecting their structure, which would show that they have different amplitudes for the different $\sigma_z$ values and no discernible symmetry in the harmonic oscillator's wave function.

At general values of $\epsilon$, the energy levels exhibit avoided crossings when plotted as functions of $\Delta$ and/or $g$, as can be seen in Fig.~\ref{Fig:EnergyLevels}(b). This result agrees with the expectation that the energy levels of systems that lack symmetries should avoid exact crossings. Instead avoided crossing structures are observed when the energy levels are plotted as functions of some system parameter.

Interestingly, when $\epsilon$ is taken close to an integer multiple of $\omega$, the energy gaps at the avoided crossings become small. When $\epsilon$ is exactly an integer multiple of $\omega$, i.e.~$\epsilon/\omega=n$ with any integer number $n$, the gaps vanish [see Figs.~\ref{Fig:EnergyLevels}(c,d)]. This point has already been established in the literature \cite{Braak2011,Li,Wakayama,Kimoto}. Our own numerical calculations [e.g.~as shown in Fig.~\ref{Fig:EnergyLevels}(d)] show that if there are gaps at the crossings, they would be upper-bounded by a scale that is at least eight orders of magnitude smaller than the other system parameters. Taking into consideration that these extremely small numerical values are obtained even when all the other parameters are comparable (but not too close) to each other, the only seemingly reasonable explanation for these numerical results is that the energy levels are indeed exhibiting crossings and not avoided crossings with small gaps. Besides, the fact that there is a large number of these crossings with a steady pattern indicates that the crossings cannot be attributed to a coincidence related to the values of the chosen parameters. The existence of these energy level crossings then suggests that there is a symmetry in the Hamiltonian, even if it is not seen as intuitively and clearly as in the case $\epsilon=0$. In the following three sections, we analyze various aspects of the energy levels and eigenstates and try to gain some insight into the nature of the symmetry.

\section{Searching for a symmetry-based partitioning of the Hilbert space}
\label{Sec:SearchingForPartition}

We now attempt to identify the symmetry numerically based on the following argument. Typically, when there is a symmetry in a quantum system, the full Hilbert space can be partitioned into smaller subspaces and the matrix elements of the Hamiltonian connecting states from different subspaces vanish. For example, if we consider a system composed of a collection of spins, the Hilbert space can be partitioned into subspaces, each of which has a well-defined total spin value. If the system possesses rotation symmetry, each energy eigenstate will (or at least can always be defined to) have a well-defined total spin. In other words, each energy eigenstate will be a superposition of states that all belong to the same subspace. Although the exact superposition will depend on the details of the interactions in the system, and different system Hamiltonians will generally result in different energy eigenstates, based on the symmetry alone we know that each energy eigenstate will belong to a subspace of well-defined total spin. In this sense, the different subspaces do not mix.

Another example of the symmetry-based Hilbert-space partitioning occurs in the symmetric QRM. In the case $\epsilon=g=0$, the energy eigenstates can be divided into two subspaces that together span the full Hilbert space:
\begin{eqnarray}
\mathcal{H}_+ & = & \left\{ \ket{\downarrow,0}, \ket{\uparrow,1} , \ket{\downarrow,2}, ... \right\}, \nonumber \\
\mathcal{H}_- & = & \left\{ \ket{\uparrow,0}, \ket{\downarrow,1} , \ket{\uparrow,2}, ... \right\},
\label{Eq:SymmetricAndAntisymmetricBasisStates}
\end{eqnarray}
where $\ket{\uparrow}$ and $\ket{\downarrow}$ are the eigenstates of the qubit operator $\hat{\sigma}_z$ (with $\hat{\sigma}_z\ket{\uparrow}=\ket{\uparrow}$ and $\hat{\sigma}_z\ket{\downarrow}=-\ket{\downarrow}$), and the second index in the ket represents the number of excitation quanta in the harmonic oscillator. If we now take the Hamiltonian $\hat{H}$ for any values of $\Delta$ and $g$ (keeping $\epsilon=0$), and we select any pair of states $\ket{\psi_+}\in\mathcal{H}_+$ and $\ket{\psi_-}\in\mathcal{H}_-$, we find that
\begin{equation}
\bra{\psi_+} \hat{H} \ket{\psi_-} = 0.
\end{equation}
In other words, the Hamiltonian does not mix states that belong to different subspaces. A consequence of this result is that for any values of $\Delta$ and $g$ each energy eigenstate of the Hamiltonian will be a superposition of states exclusively in $\mathcal{H}_+$ or exclusively in $\mathcal{H}_-$. (In cases of degeneracy the energy eigenstates do not have to, but can always be chosen to, satisfy the above statement.) The two subspaces do not mix. They are spaces of different symmetry, which in this case is the parity, and the Hamiltonian does not change the parity of a state it operates on. As a further consequence, if we take a positive-parity energy eigenstate $\ket{\psi_+(0,\Delta_1/\omega,g_1/\omega)}$ obtained for parameters $\epsilon=0$, $\Delta_1/\omega$ and $g_1/\omega$ and a negative-parity energy eigenstate $\ket{\psi_-(0,\Delta_2/\omega,g_2/\omega)}$ obtained for parameters $\epsilon=0$, $\Delta_2/\omega$ and $g_2/\omega$, we find that
\begin{equation}
\braket{\psi_+(0,\Delta_1/\omega,g_1/\omega)}{\psi_-(0,\Delta_2/\omega,g_2/\omega)} = 0.
\label{Eq:ZeroOverlap}
\end{equation}

Another, perhaps simpler, way to derive Eq.~(\ref{Eq:ZeroOverlap}) is to say that if $\ket{\psi_1}$ and $\ket{\psi_2}$ are two eigenstates of a symmetry operator $\hat{S}$,
\begin{eqnarray}
\hat{S} \ket{\psi_1} & = & S_1 \ket{\psi_1}, \nonumber \\
\hat{S} \ket{\psi_2} & = & S_2 \ket{\psi_2},
\end{eqnarray}
with different eigenvalues $S_1$ and $S_2$, then we have the relations
\begin{eqnarray}
\bra{\psi_1} \hat{S} \ket{\psi_2} & = & S_1 \braket{\psi_1}{\psi_2}, \nonumber \\
\bra{\psi_1} \hat{S} \ket{\psi_2} & = & S_2 \braket{\psi_1}{\psi_2}.
\end{eqnarray}
These two equations cannot be simultaneously satisfied unless $\braket{\psi_1}{\psi_2} = 0$, which corresponds to Eq.~(\ref{Eq:ZeroOverlap}) in the present case.

We can then try to identify two, or possibly a few, subspaces that give the same result in the case $\epsilon/\omega=n$ where $n$ is a finite integer. Let us assume that we have a given value of $n$. In order to identify subspaces that the Hamiltonian does not mix, we start by setting $\Delta$ and $g$ to some arbitrary values $\Delta_1$ and $g_1$. To avoid unrelated degeneracies in the spectrum that could create unnecessary complications in the calculations, we specifically avoid the special case $g=0$. We diagonalize the Hamiltonian with the chosen parameters. Thus we obtain a basis $\mathcal{B}_1$ that spans the full Hilbert space. The idea now is to search for a partitioning of $\mathcal{B}_1$ into two (or more) sets that the Hamiltonian does not mix, regardless of the values of $\Delta_1$ and/or $g_1$. For this purpose we choose different values of $\Delta$ and/or $g$ (to which we refer as $\Delta_2$ and $g_2$), and we diagonalize the Hamiltonian to find another basis set $\mathcal{B}_2$. If we take the overlap between each state in $\mathcal{B}_1$ and all the states in $\mathcal{B}_2$, we would hope that at least for the lowest few hundred energy levels each state in $\mathcal{B}_1$ will have exactly zero overlap with about half of the lowest energy eigenstates in $\mathcal{B}_2$, helping us identify the natural partitioning of the Hilbert space. (Here we say ``few hundred energy levels'' because the overlap will decrease as a function of basis state index and, using numerical results of finite accuracy, at some point it becomes impossible to distinguish between numerical errors and very small values of the overlap.) The idea would then be that we can choose any other values of $\Delta$ and $g$, and the partitioning of the Hilbert space would remain the same. This procedure for example works and can be used to partition the Hilbert space into two subspaces for the well-known symmetric case $\epsilon=0$, as we shall explain shortly. Establishing such a partitioning of the Hilbert space for finite values of $\epsilon/\omega$ would be helpful in the search for the hidden symmetry.

\begin{figure}[h]
\includegraphics[width=16.0cm]{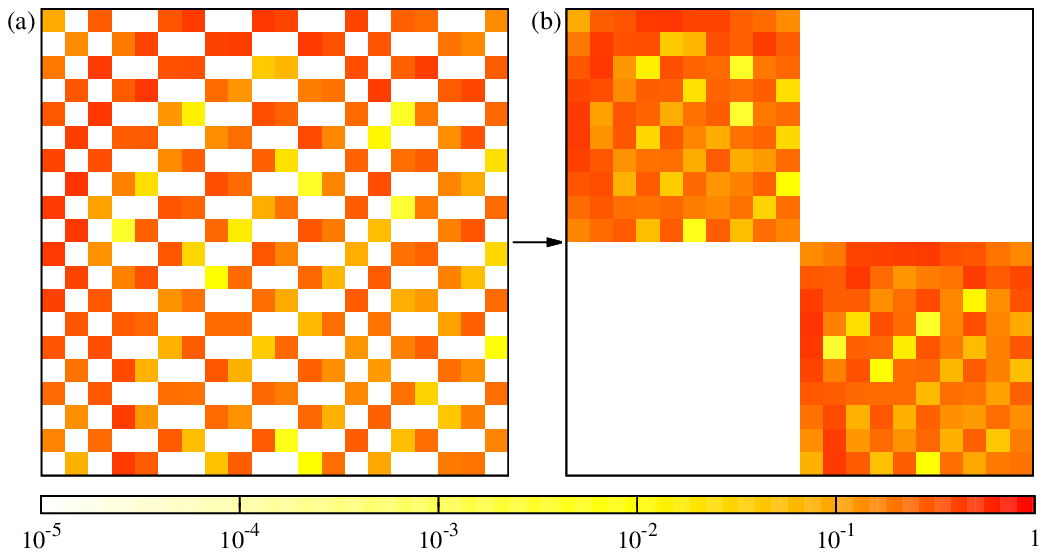}
\caption{Overlap $\left|\braket{\psi_n(0,\Delta/\omega,g_1/\omega)}{\psi_m(0,\Delta/\omega,g_2/\omega)}\right|$ between the energy eigenstates $\ket{\psi_n(0,\Delta/\omega,g/\omega)}$ obtained with the parameters $(\epsilon/\omega,\Delta/\omega,g_1/\omega)=(0,0.7,0.5)$ [with each of these states represented in one row] and those obtained with the parameters $(\epsilon/\omega,\Delta/\omega,g_2/\omega)=(0,0.7,2.6)$ [with each one of these states represented in one column]. In Panel (a) the energy eigenstates are ordered according to their energies, with the ground states represented at the top left. In Panel (b) the states are reordered so as to highlight the partitioning of the states into two groups based on their symmetry. In constructing the matrix in Panel (b), we take the rows in Panel (a) in the order (1, 3, 5, 7, 9, 11, 13, 15, 17, 19, 2, 4, 6, 8, 10, 12, 14, 16, 18, 20), and we take the columns in Panel (a) in the order (1, 3, 6, 7, 10, 11, 14, 16, 17, 20, 2, 4, 5, 8, 9, 12, 13, 15, 18, 19). Panel (b) shows clearly that the energy eigenstates can be divided into two groups with no overlap between states taken from different groups.}
\label{Fig:OverlapMatrix0}
\end{figure}

First, as a demonstration, we consider the symmetric case $\epsilon=0$. We set $(\Delta/\omega,g_1/\omega)=(0.7,0.5)$ and $(\Delta/\omega,g_2/\omega)=(0.7,2.6)$, and we obtain two sets of energy eigenstates. We then take the lowest 20 energy levels from each eigenbasis and calculate the overlaps $\left|\braket{\psi_n(0,\Delta/\omega,g_1/\omega)}{\psi_m(0,\Delta/\omega,g_2/\omega)}\right|$ between all the different combinations of energy eigenstates taken from different eigenbases. The results are shown in Fig.~\ref{Fig:OverlapMatrix0} and Table \ref{Tab:OverlapMatrix0}. The figure and table show clearly that each eigenbasis can be divided into two groups, such that the states in each group in $\mathcal{B}_1$ have zero overlap with the states in one group in $\mathcal{B}_2$. The reason is of course that each state in $\mathcal{B}_1$ or in $\mathcal{B}_2$ has either positive or negative parity, and the overlap between states of different parity must vanish.

\begin{figure}[h]
\includegraphics[width=16.0cm]{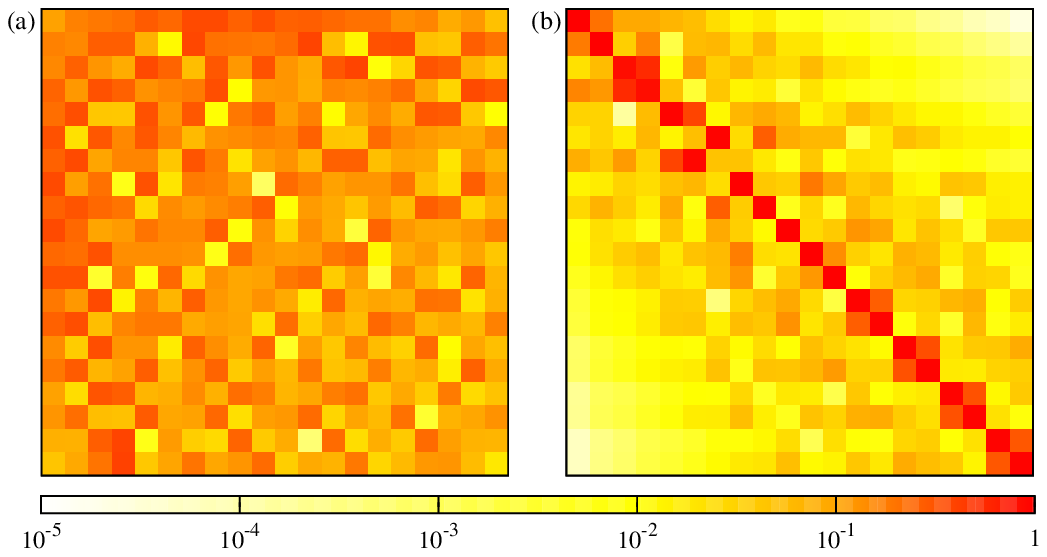}
\caption{Overlap $\left|\braket{\psi_n(1,\Delta_1/\omega,g_1/\omega)}{\psi_m(1,\Delta_2/\omega,g_2/\omega)}\right|$ between the energy eigenstates $\ket{\psi_n(1,\Delta/\omega,g/\omega)}$ obtained with the parameters $(\epsilon/\omega,\Delta_1/\omega,g_1/\omega)$ and those obtained with the parameters $(\epsilon/\omega,\Delta_2/\omega,g_1/\omega)$. In Panel (a) we set $(\epsilon/\omega,\Delta_1/\omega,g_1/\omega)=(1,0.7,0.5)$ and $(\epsilon/\omega,\Delta_2/\omega,g_2/\omega)=(1,0.7,2.6)$. In Panel (b) we set $(\epsilon/\omega,\Delta_1/\omega,g_1/\omega)=(1,0.7,0.5)$ and $(\epsilon/\omega,\Delta_2/\omega,g_2/\omega)=(1,1.8,0.5)$. It is clear that neither case allows a grouping of the energy eigenstates that would result in vanishing overlaps between members of the different groups.}
\label{Fig:OverlapMatrix1}
\end{figure}

We now follow the same procedure for the case $\epsilon/\omega=1$. The results are shown in Fig.~\ref{Fig:OverlapMatrix1} and Tables \ref{Tab:OverlapMatrix1VaryG} and \ref{Tab:OverlapMatrix1VaryDelta}. We show results in which we take a pair of parameter sets differing in the $g$ value and a pair of parameter sets differing in the $\Delta$ value. Clearly there are no zeroes in any of these results. It is therefore impossible to partition each energy eigenbasis into groups in such a way that certain groups of states have no overlap with groups of states from other eigenbases (obtained using different values of $\Delta$ and/or $g$).

It is perhaps worth noting here that the absence of zero overlaps remains mostly true even if we choose parameter sets that correspond to energy level crossings. One complication in such a case is that the degenerate energy eigenstates can be redefined to produce alternative energy eigenstates, because any linear superposition of states that have the same energy will also have the same energy, and one can always define the basis so as to eliminate the overlap with any desired energy eigenstate taken from another eigenbasis. However, all other overlaps will in general remain finite.

The impossibility of grouping the energy eigenstates in such a way that the overlap between an energy eigenstate (which would have a certain symmetry value) taken from the eigenbasis $\mathcal{B}_1$ and an energy eigenstate of a different symmetry value taken from the eigenbasis $\mathcal{B}_2$ has important implications about the symmetry in the system. For example, if there is a symmetry operator $\hat{S}$ such that
\begin{equation}
\hat{S} \ket{\psi_n} = S_+ \ket{\psi_n} \ \ {\rm or} \ \ \hat{S} \ket{\psi_n} = S_- \ket{\psi_n},
\end{equation}
where for simplicity in appearance we have assumed that there are only two symmetry values ($S_+$ and $S_-$), then it must be the case that the operator $\hat{S}$ depends on the system parameters $\Delta/\omega$ and $g/\omega$. This situation stands in stark contrast to conventional symmetries where the symmetry operator is expected to be independent of system parameters and should only reflect general properties about the eigenstates of the symmetry operator. An example is the parity operator given in Eq.~(\ref{Eq:ParityOperator}).

\section{Discussion}
\label{Sec:Discussion}
 
In retrospect, one could say that it is not surprising that the symmetry operator cannot be independent of system parameters in the case $\epsilon\neq 0$. This point can be seen by setting $\epsilon=g=0$ and varying $\Delta/\omega$. The energy eigenstates in this case are given by Eq.~(\ref{Eq:SymmetricAndAntisymmetricBasisStates}), which means that the energy eigenstates are independent of $\Delta/\omega$. When $\epsilon$ is finite (e.g.~when $\epsilon/\omega=1$), the energy eigenstates (even when $g=0$) will depend on $\Delta/\omega$. Specifically, the qubit Hamiltonian eigenstates transform gradually from being eigenstates of $\hat{\sigma}_z$ to being eigenstates of $\hat{\sigma}_x$ as $\Delta/\epsilon$ is increased from 0 to $\infty$. The fact that the qubit Hamiltonian eigenstates continuously transform and essentially cover the full range of qubit bases means that there cannot be any simple symmetry classification of the states that is independent of system parameters.

One could also say that, even though we have established that the symmetry operator cannot be written in simple, parameter-independent form, such an operator could still be defined formally once the different relevant subspaces (i.e.~the groups of quantum states divided based on their symmetry) are identified. These subspaces can in fact be identified by inspecting the energy level structure. For example, by looking at Fig.~\ref{Fig:EnergyLevels}(c), one can say that any two energy levels that cross each other almost certainly correspond to quantum states that belong to different symmetry groups. Conversely, energy levels that exhibit avoided-crossing structures belong to the same group. One example of such a situation is the structures that look like typical avoided level crossings in the range $0.2<g/\omega<0.4$ in Fig.~\ref{Fig:EnergyLevels}(c). Thus the entire Hilbert space can be partitioned into subspaces
\begin{equation}
\mathcal{H}_i (\Delta/\omega,g/\omega) = \{ \ket{\psi_{i,1}(\Delta/\omega,g/\omega)}, \ket{\psi_{i,2}(\Delta/\omega,g/\omega)}, ... \}.
\end{equation}
The two indices in the subscript of $\psi$ can be understood as the quantum numbers that specify the corresponding quantum state. Importantly, all the states with the same value of $i$ have the same value of the symmetry operator. As a result, once the partitioning is established, projection operators can be defined,
\begin{equation}
\hat{P}_i (\Delta/\omega,g/\omega) = \sum_j \ket{\psi_{i,j}(\Delta/\omega,g/\omega)}\bra{\psi_{i,j}(\Delta/\omega,g/\omega)},
\end{equation}
from which the symmetry operator can be defined:
\begin{equation}
\hat{S} (\Delta/\omega,g/\omega) = \sum_i S_i \hat{P}_i (\Delta/\omega,g/\omega),
\end{equation}
where $S_i$ can be chosen as any set of distinct numbers that distinguish the states based on their symmetry properties. However, since we do not see how this exercise will lead us to a simple definition of the symmetry operator (i.e.~expressed as a relatively simple function of the basic qubit and harmonic oscillator operators) or insight into the nature of the symmetry, we shall not pursue this calculation further.

\begin{figure}[h]
\includegraphics[width=8.0cm]{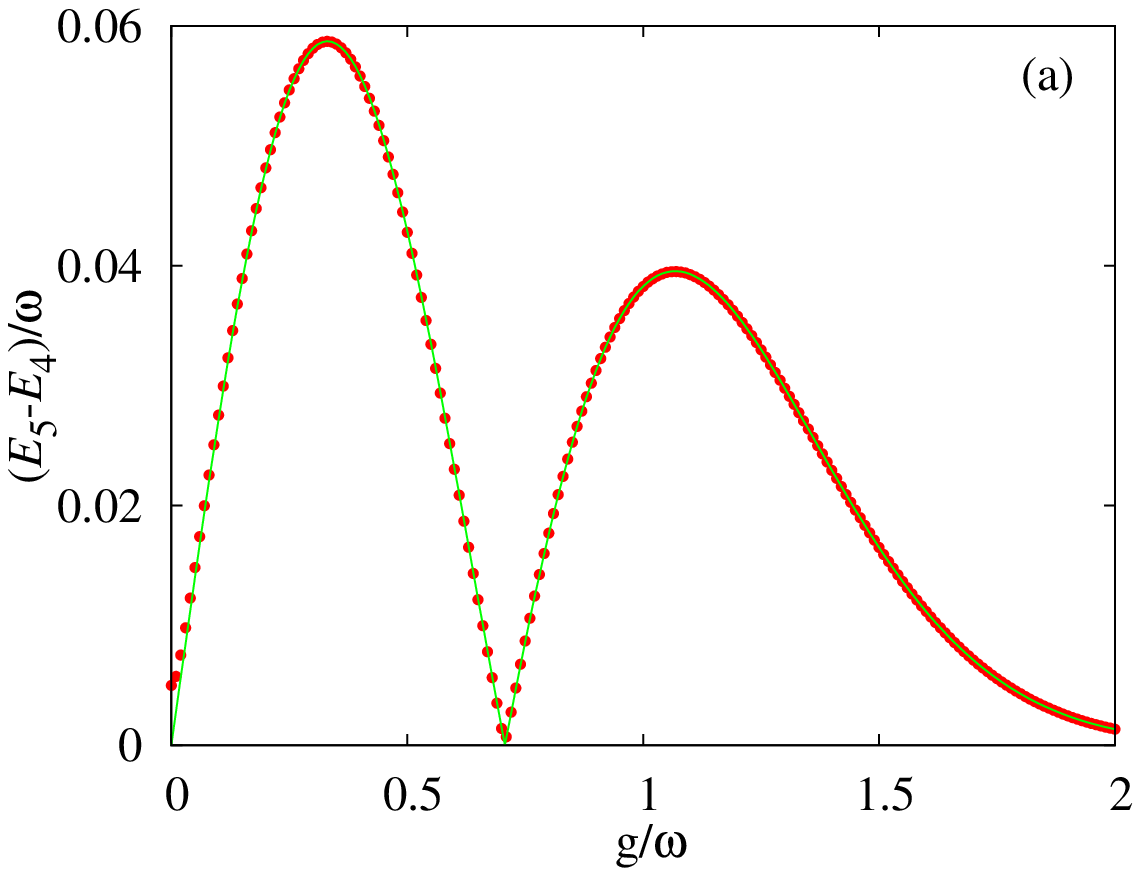}
\includegraphics[width=8.0cm]{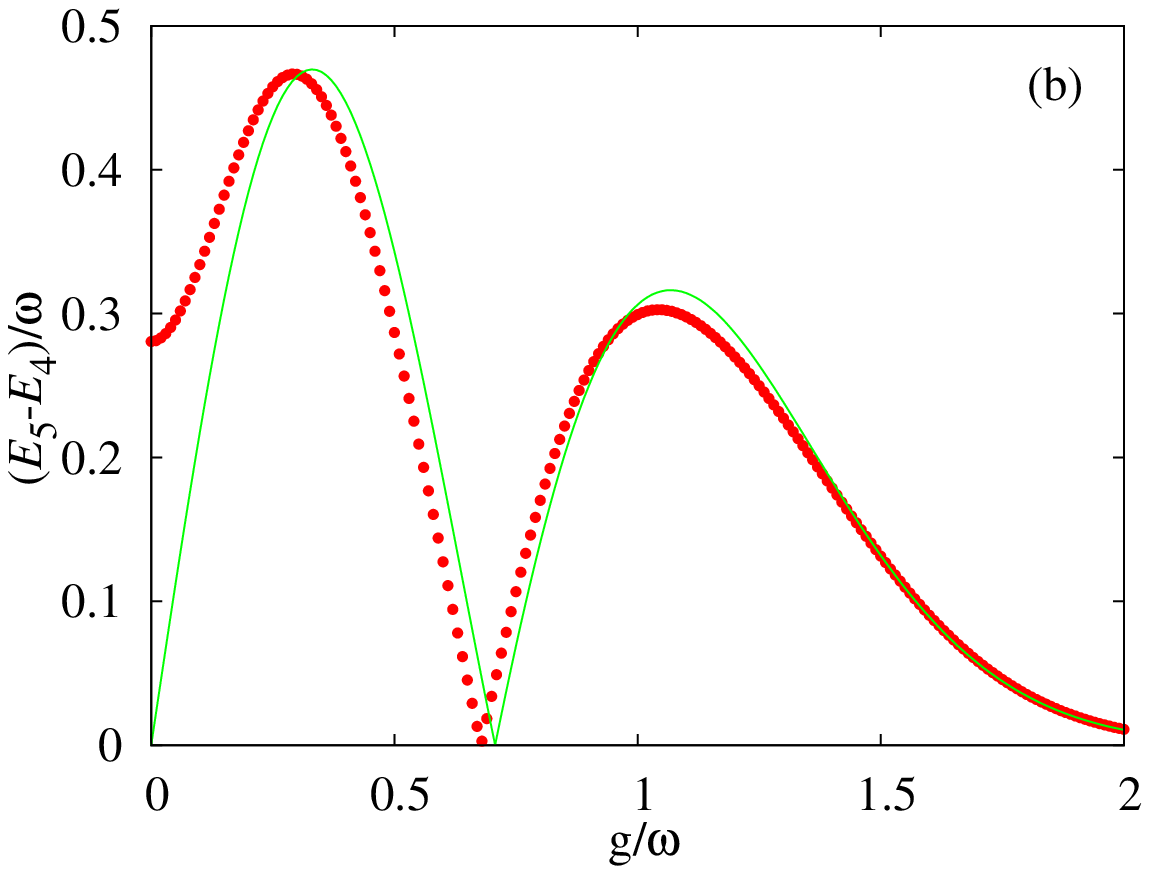}
\caption{The difference between energy levels number 4 and 5 (measured in units of $\omega$) as a function of coupling strength $g/\omega$ for $\epsilon/\omega=1$. In Panel (a) $\Delta/\omega=0.1$, while in Panel (b) $\Delta/\omega=0.8$. The dots are obtained numerically by diagonalizing the Hamiltonian and are therefore essentially exact. The solid line is given by the analytical formula based on Eq.~(\ref{Eq:PerturbativeOffDiagonalMatrixElement}), i.e.~$E_5-E_4=2^{-1/2}\Delta\exp\{-2g^2/\omega^2\}(2g/\omega)L_1^1[(2g/\omega)^2]$.}
\label{Fig:EnergyLevelSplitting}
\end{figure}

We now consider what we can say about the present problem from the point of view of the rotating-wave approximation of Ref.~\cite{Irish}. In the limit where $\Delta/\omega$ is small or $g/\omega$ is large, there is a generalized rotating-wave approximation that works well when $\epsilon\approx n\omega$ (including the case of exact equality $\epsilon=n\omega$) \cite{Ashhab2010}, which is related to the perturbation-theory approach of Refs.~\cite{Larson,Semple}. Under this approximation we can treat the term $\Delta\hat{\sigma}_z /2$ in the Hamiltonian as a perturbation. If we ignore this perturbation, we can easily see that the energy eigenstates are given by $\ket{\leftarrow}\otimes\hat{D}(\alpha)\ket{m}$ and $\ket{\rightarrow}\otimes\hat{D}(-\alpha)\ket{m}$, where $\ket{\leftarrow}$ and $\ket{\rightarrow}$ are eigenstates of $\hat{\sigma}_x$ (with $\hat{\sigma}_x\ket{\rightarrow}=\ket{\rightarrow}$ and $\hat{\sigma}_x\ket{\leftarrow}=-\ket{\leftarrow}$) and $\hat{D}$ is a displacement operator that transforms the harmonic oscillator's Fock states $\ket{m}$ such that they are shifted in the positive or negative $x$ direction to account for the qubit-state dependent (but otherwise constant) force imparted by the qubit on the oscillator [$\hat{D}(\alpha)=\exp\{\alpha\hat{a}^{\dagger}-\alpha^*\hat{a}\}$]. The displacement is given by $\alpha=g/\omega$. The energies of these states in the absence of the perturbation are given by $\pm\epsilon/2 + n\omega$, up to an overall constant. The Hilbert space can therefore be partitioned into small subspaces that are one dimensional for the lowest $n$ energy levels and two dimensional for the higher energy levels (i.e.~forming energy level pairs), with each such subspace being well separated in energy from the rest of the Hilbert space. The perturbation $\Delta\hat{\sigma}_z /2$ will then mix the states within each two-dimensional subspace, but it will cause little mixing between states from different subspaces. Within a single two-dimensional subspace, the effect of the perturbation can be calculated using the effective Hamiltonian
\begin{equation}
\hat{H}_{\rm eff} = \frac{1}{2} \left( \begin{array}{cc}
\epsilon - n\omega & \tilde{\Delta}_{mn} \\
\tilde{\Delta}_{mn} & -(\epsilon - n\omega)
\end{array}
\right) + {\rm constant},
\label{Eq:2DEffectiveHamiltonian}
\end{equation}
with the off-diagonal matrix elements
\begin{eqnarray}
\tilde{\Delta}_{mn} & = & \Delta \bra{\leftarrow}\otimes\bra{m}\hat{D}(-g/\omega)\hat{\sigma}_z\hat{D}(-g/\omega)\ket{\rightarrow}\otimes\ket{m-n} \nonumber \\
& = & \Delta \bra{m}\hat{D}(-2g/\omega)\ket{m-n}.
\end{eqnarray}
These matrix elements (for any integer $m$ with $m\geq n$) are given by \cite{Ashhab2010}:
\begin{equation}
\tilde{\Delta}_{mn} = \Delta e^{-2g^2/\omega^2} \left( -\frac{2g}{\omega} \right)^n \sqrt{\frac{(m-n)!}{m!}} L_{m-n}^{n}\left[ \frac{4g^2}{\omega^2} \right],
\label{Eq:PerturbativeOffDiagonalMatrixElement}
\end{equation}
where $L_i^j(x)$ are the associated Laguerre polynomials. The main properties of $\tilde{\Delta}_{mn}$ are as follows: they vanish in the special case $g=0$ and $n\neq 0$, they vanish asymptotically when $g/\omega\rightarrow\infty$, and they change sign $m-n$ times at finite values of $g$, giving rise to $m-n$ energy crossing points. The effect of coupling to states outside the two-dimensional subspace under consideration can be included using Van Vleck perturbation theory, as explained e.g.~in Ref.~\cite{Ashhab2008}. This coupling will modify the values of the matrix elements in Eq.~(\ref{Eq:2DEffectiveHamiltonian}). Importantly, however, because the energy crossing points at finite values of $g$ occur at points where both the diagonal and off-diagonal matrix elements in $\hat{H}_{\rm eff}$ change sign as functions of the system parameters, small corrections to the matrix elements can shift the locations of the crossing points but will not affect the existence of these crossings.

Two examples of the energy level separation between two neighboring energy levels as a function of $g$ are plotted in Fig.~\ref{Fig:EnergyLevelSplitting}. The figure shows that the perturbation-theory approximation works very well (and becomes exact) in the limits $\Delta/\omega\rightarrow 0$ or $g/\omega\rightarrow\infty$, as mentioned above. Deviations from the theoretical formula, which result from coupling to states outside the two-dimensional subspace, become clearly visible away from these limits. Nevertheless, the existence of the energy-level crossing is not affected by the perturbation, even when this perturbation is quite large [Fig.~\ref{Fig:EnergyLevelSplitting}(b)].

The perturbation-theory analysis also highlights the fact that the quantum states of the unperturbed basis (which capture the symmetry in certain limits) depend on the system parameters (because the displacement $\alpha=g/\omega$), which suggests that the symmetry in this problem involves basis states that depend on the system parameters, unlike the usual case where the symmetry is defined based on the overall form of the terms in the Hamiltonian but is independent of the values of the system parameters.

Before concluding this section, we make a final observation about the degeneracies in the spectrum. The proof of the existence of degeneracies was made possible because the level crossings correspond to quasi-exact (Juddian) solutions, just as the degenerate states in the symmetric QRM. They are therefore always located on the baselines of the model with energies $E_b(n) = n\omega-g^2/\omega \pm \epsilon/2$, i.e.~the exceptional spectrum. One can prove that the asymmetric QRM cannot possess any degeneracies in its regular spectrum \cite{Braak2011,Braak2019}. Therefore the search for possible degeneracies away from integer values of $\epsilon/\omega$ could be restricted to the baselines. These facts may turn out to be related to the hidden symmetry operator (if it exists).

\section{Searching for additional points with hidden symmetry}
\label{Sec:NonIntegerEpsilonValues}

\begin{figure}[h]
\includegraphics[width=8.0cm]{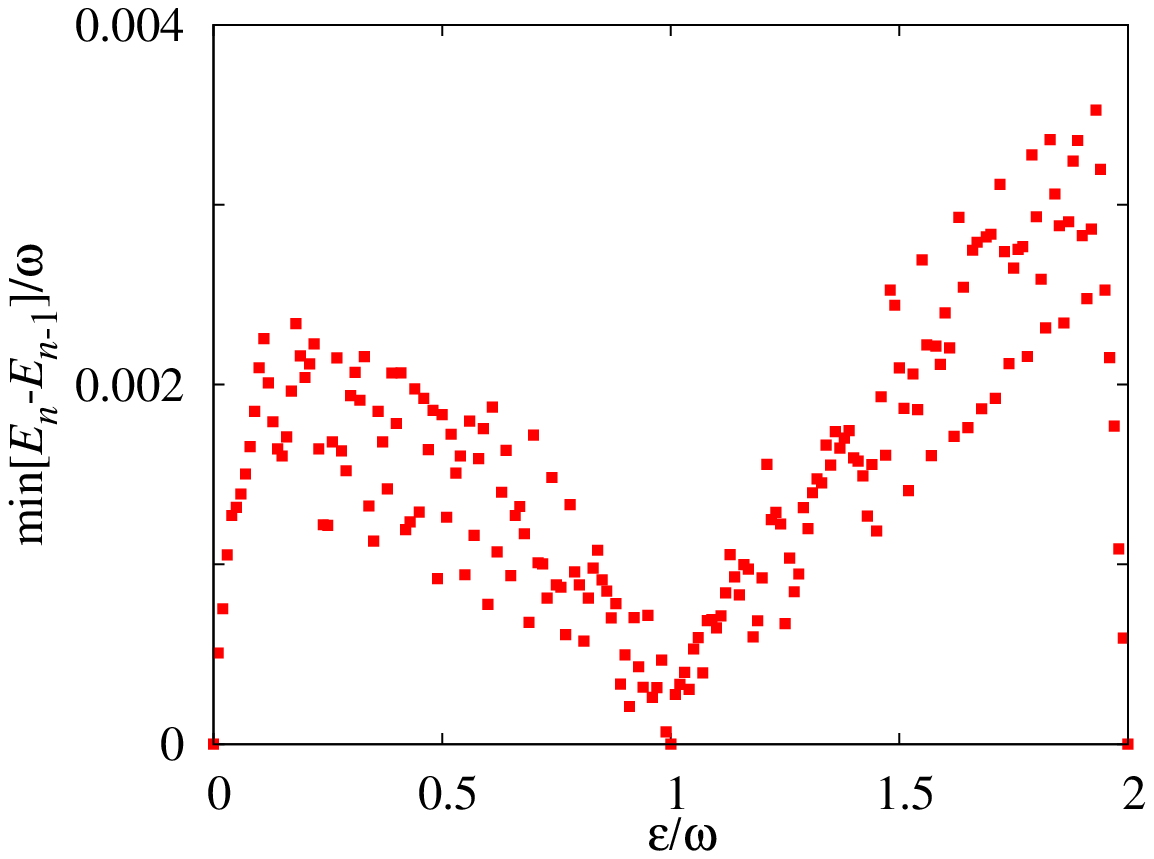}
\caption{Minimum energy gap between any two energy levels within the ten lowest levels as a function of $\epsilon/\omega$. The minimum gap is taken over $(\Delta/\omega,g/\omega)$ values in the range $\Delta/\omega\in [0.1,3.1]$ and $g/\omega\in [0.1,3.1]$ with a mesh of spacing 0.01 between neighboring points in $(\Delta/\omega,g/\omega)$ parameter space. The minimum gap is effectively zero at integer values of $\epsilon/\omega$ but not at any other value of $\epsilon/\omega$.}
\label{Fig:MinimumGapAsFunctionOfBias}
\end{figure}

So far we have focused on the case where $\epsilon/\omega$ is an integer, where it has already been established that energy level crossings exist. In this section we raise the question of whether there could be other cases with non-integer values of $\epsilon/\omega$ where energy level crossings exist. For this purpose, we scan the parameters $(\epsilon/\omega,\Delta/\omega,g/\omega)$ in the ranges $\epsilon/\omega\in [0,2]$, $\Delta/\omega\in [0.1,3.1]$ and $g/\omega\in [0.1,3.1]$ and look for possible energy level crossings. For each value of $\epsilon/\omega$, we use a mesh of $301 \times 301$ points in the $(\Delta/\omega,g/\omega)$ space, i.e.~using a spacing of 0.01 in the parameter values, and we calculate the minimum energy gap within the lowest ten energy levels. In Fig.~\ref{Fig:MinimumGapAsFunctionOfBias}, we plot the minimum gap as a function of $\epsilon/\omega$, i.e.~taking the minimum over all possible values of $\Delta/\omega$ and $g/\omega$. As expected, the minimum gap vanishes at integer values of $\epsilon/\omega$. We find no additional values of $\epsilon/\omega$ that result in extremely small energy gaps that could serve as signs of energy level crossings. We therefore conclude, with a high degree of confidence, that there are no additional cases of hidden symmetry in the asymmetric QRM.

\section{Conclusion}
\label{Sec:Conclusion}

We have performed numerical calculations that confirm the existence of energy level crossings in the asymmetric QRM at integer values of $\epsilon/\omega$. We have further made an attempt at identifying the hidden symmetry in the asymmetric QRM. Our results lead to the conclusion that the symmetry operator must depend on the different system parameters, which means that it cannot have a simple intuitive interpretation as for example in the case of the symmetric QRM. We have also searched and concluded that it is highly unlikely that there are additional cases of hidden symmetry beyond what has been established in the literature. Our results shed light on the nature of the hidden symmetry and could help in the eventual identification of its full nature.

We would like to thank D. Braak, K. Semba, M. Wakayama and F. Yoshihara for useful discussions.

\section*{Appendix: values for the eigenbasis overlap matrices}

In this appendix we show the values of the overlap matrices plotted in Figs.~\ref{Fig:OverlapMatrix0} and \ref{Fig:OverlapMatrix1}.

\begin{widetext}

\begin{table}[h]
\caption{Overlap $\left|\braket{\psi_n(0,\Delta/\omega,g_1/\omega)}{\psi_m(0,\Delta/\omega,g_2/\omega)}\right|$ between the energy eigenstates $\ket{\psi_n(\epsilon/\omega,\Delta/\omega,g/\omega)}$ obtained with the parameters $(\epsilon/\omega,\Delta/\omega,g_1/\omega)=(0,0.7,0.5)$ [with each of these states represented in one row] and those obtained with the parameters $(\epsilon/\omega,\Delta/\omega,g_2/\omega)=(0,0.7,2.6)$ [with each one of these states represented in one column]. The top row and left-most column contain the state index based on the energy, starting from the ground state and going up in energy. In the tables we keep the lowest ten energy eigenstates in each eigenbasis. In this table, entries given as ``0'' indicate that the numerically obtained values were below $10^{-8}$ and are hence within the numerical error range; in the data shown in this table and in Fig.~\ref{Fig:OverlapMatrix0}, the largest of the ignored values was $1.6\times 10^{-9}$, while the smallest of the retained values was $2.6\times 10^{-3}$, meaning that there is a clear separation in scale between the finite values and the essentially vanishing values.}
\label{Tab:OverlapMatrix0}
\begin{tabular}{|c|c|c|c|c|c|c|c|c|c|c|}
\hline
  & 1 & 2 & 3 & 4 & 5 & 6 & 7 & 8 & 9 & 10 \\ \hline
1 & 0.085872 & 0 & 0.270998 & 0 & 0 & 0.29969 & 0.436304 & 0 & 0 & 0.454386 \\ \hline
2 & 0 & 0.14456 & 0 & 0.173241 & 0.382144 & 0 & 0 & 0.406689 & 0.424614 & 0 \\ \hline
3 & 0.188254 & 0 & 0.439777 & 0 & 0 & 0.313363 & 0.32688 & 0 & 0 & 0.03759 \\ \hline
4 & 0 & 0.291976 & 0 & 0.272117 & 0.457068 & 0 & 0 & 0.226131 & 0.123849 & 0 \\ \hline
5 & 0.293962 & 0 & 0.461887 & 0 & 0 & 0.116467 & 0.0143253 & 0 & 0 & 0.328976 \\ \hline
6 & 0 & 0.412869 & 0 & 0.26795 & 0.274358 & 0 & 0 & 0.13769 & 0.210023 & 0 \\ \hline
7 & 0.378705 & 0 & 0.329382 & 0 & 0 & 0.140276 & 0.259821 & 0 & 0 & 0.255201 \\ \hline
8 & 0 & 0.469615 & 0 & 0.160232 & 0.0225654 & 0 & 0 & 0.330362 & 0.219467 & 0 \\ \hline
9 & 0.427926 & 0 & 0.103675 & 0 & 0 & 0.298293 & 0.243356 & 0 & 0 & 0.0772449 \\ \hline
10 & 0 & 0.452834 & 0 & 0.00271208 & 0.251535 & 0 & 0 & 0.236954 & 0.0154019 & 0 \\ \hline
\end{tabular}
\end{table}

\begin{table}[h]
\caption{Overlap $\left|\braket{\psi_n(1,\Delta/\omega,g_1/\omega)}{\psi_m(1,\Delta/\omega,g_2/\omega)}\right|$ between the energy eigenstates $\ket{\psi_n(\epsilon/\omega,\Delta/\omega,g/\omega)}$ obtained with the parameters $(\epsilon/\omega,\Delta/\omega,g_1/\omega)=(1,0.7,0.5)$ and those obtained with the parameters $(\epsilon/\omega,\Delta/\omega,g_2/\omega)=(1,0.7,2.6)$.} 
\label{Tab:OverlapMatrix1VaryG}
\begin{tabular}{|c|c|c|c|c|c|c|c|c|c|c|}
\hline
  & 1 & 2 & 3 & 4 & 5 & 6 & 7 & 8 & 9 & 10 \\ \hline
1 & 0.102807 & 0.165425 & 0.18125 & 0.204599 & 0.28147 & 0.231363 & 0.340805 & 0.354332 & 0.24836 & 0.325739\\  \hline
2 & 0.160805 & 0.150032 & 0.263796 & 0.266683 & 0.0713559 & 0.0114136 & 0.357451 & 0.200113 & 0.192774 & 0.188331 \\ \hline
3 & 0.14731 & 0.250945 & 0.123814 & 0.0842062 & 0.355356 & 0.242037 & 0.0532981 & 0.290481 & 0.118781 & 0.319978 \\ \hline
4 & 0.247986 & 0.119294 & 0.321285 & 0.25689 & 0.130891 & 0.152198 & 0.178257 & 0.333624 & 0.00977145 & 0.115952 \\ \hline
5 & 0.214312 & 0.336138 & 0.0412139 & 0.0397512 & 0.313538 & 0.122879 & 0.287124 & 0.00929818 & 0.182471 & 0.249432 \\ \hline
6 & 0.314164 & 0.0217168 & 0.294536 & 0.148904 & 0.292471 & 0.15421 & 0.0567743 & 0.126906 & 0.138425 & 0.16277 \\ \hline
7 & 0.253239 & 0.351607 & 0.0978355 & 0.157374 & 0.151457 & 0.0427969 & 0.30267 & 0.176537 & 0.0202042 & 0.105306 \\ \hline
8 & 0.348607 & 0.109879 & 0.206735 & 0.00525292 & 0.315546 & 0.01923 & 0.179177 & 0.164169 & 0.10805 & 0.000722655 \\ \hline
9 & 0.256235 & 0.301374 & 0.23521 & 0.195111 & 0.0254403 & 0.140025 & 0.111149 & 0.143323 & 0.172429 & 0.271095 \\ \hline
10 & 0.350104 & 0.234847 & 0.097638 & 0.112868 & 0.193846 & 0.149183 & 0.147824 & 0.269674 & 0.00748982 & 0.13179 \\ \hline
\end{tabular}
\end{table}

\begin{table}[h]
\caption{Overlap $\left|\braket{\psi_n(1,\Delta_1/\omega,g/\omega)}{\psi_m(1,\Delta_2/\omega,g/\omega)}\right|$ between the energy eigenstates $\ket{\psi_n(\epsilon/\omega,\Delta/\omega,g/\omega)}$ obtained with the parameters $(\epsilon/\omega,\Delta_1/\omega,g/\omega)=(1,0.7,0.5)$ and those obtained with the parameters $(\epsilon/\omega,\Delta_2/\omega,g/\omega)=(1,1.8,0.5)$.} 
\label{Tab:OverlapMatrix1VaryDelta}
\begin{tabular}{|c|c|c|c|c|c|c|c|c|c|c|}
\hline
  & 1 & 2 & 3 & 4 & 5 & 6 & 7 & 8 & 9 & 10 \\ \hline
1 & 0.96738 & 0.199033 & 0.0915272 & 0.0903519 & 0.0648127 & 0.05268 & 0.0232236 & 0.0121967 & 0.0168617 & 0.0047086 \\ \hline
2 & 0.175533 & 0.966042 & 0.0328179 & 0.151747 & 0.0015081 & 0.0538998 & 0.0636757 & 0.0246696 & 0.0582072 & 0.0190485 \\ \hline
3 & 0.0216759 & 0.0541826 & 0.813775 & 0.55671 & 0.0071516 & 0.115413 & 0.0358231 & 0.0618321 & 0.0311963 & 0.0239472 \\ \hline
4 & 0.156636 & 0.119545 & 0.557076 & 0.803544 & 0.047923 & 0.00235764 & 0.0423296 & 0.012997 & 0.0160923 & 0.00222915 \\ \hline
5 & 0.0303449 & 0.0312773 & 0.000140632 & 0.0550665 & 0.914144 & 0.37283 & 0.0118446 & 0.06253 & 0.0853124 & 0.0614312 \\ \hline
6 & 0.018286 & 0.0309794 & 0.0152981 & 0.061755 & 0.0130238 & 0.0492092 & 0.949661 & 0.0243169 & 0.265606 & 0.0830394 \\ \hline
7 & 0.077898 & 0.042829 & 0.114335 & 0.0312155 & 0.376124 & 0.910667 & 0.0468325 & 0.0451187 & 0.018187 & 0.00643401 \\ \hline
8 & 0.0129774 & 0.0160855 & 0.0304578 & 0.0233544 & 0.0480622 & 0.0628637 & 0.0235812 & 0.966699 & 0.0394558 & 0.0329615 \\ \hline
9 & 0.0264194 & 0.0680246 & 0.036279 & 0.0161135 & 0.0891112 & 0.00691535 & 0.264711 & 0.036906 & 0.952123 & 0.00493437 \\ \hline
10 & 0.00783376 & 0.0225809 & 0.016727 & 0.00598152 & 0.0443078 & 0.0125345 & 0.0860162 & 0.0332038 & 0.0115349 & 0.976888 \\ \hline
\end{tabular}
\end{table}

\end{widetext}


\begin{thebibliography}{99}

\bibitem{Gross} D. J. Gross, Proc. Nat. Acad. Sci. {\bf 93}, 14256 (1996).

\bibitem{Noether} E. Noether, Nachr. D. K\"onig. Gesellsch. D. Wiss. Zu G\"ottingen, Math-phys. Klasse. 235 (1918).

\bibitem{Kuklov} A. B. Kuklov and B. V. Svistunov, Phys. Rev. Lett. {\bf 89}, 170403 (2002).

\bibitem{Ashhab2003} S. Ashhab and A. J. Leggett, Phys. Rev. A {\bf 68}, 063612 (2003).

\bibitem{BooksOnSymmetryAndCrossings} See e.g.~L. D. Landau and L. M. Lifshitz, {\it Quantum Mechanics: Non-Relativistic Theory} (Butterworth-Heinemann, 1981); J. J. Sakurai, {\it Modern Quantum Mechanics} (Addison-Wesley, Reading, 1994).

\bibitem{Irish} E. K. Irish, Phys. Rev. Lett. {\bf 99}, 173601 (2007).

\bibitem{Niemczyk} T. Niemczyk, F. Deppe, H. Huebl, E. P. Menzel, F. Hocke, M. J. Schwarz, J. J. Garcia-Ripoll, D. Zueco, T. H\"ummer, E. Solano, A. Marx, and R. Gross, Nature Phys. {\bf 6}, 772 (2010).

\bibitem{Forn2010} P. Forn-Diaz, J. Lisenfeld, D. Marcos, J. J. Garcia-Ripoll, E. Solano, C. J. P. M. Harmans, and J. E. Mooij, Phys. Rev. Lett. {\bf 105}, 237001 (2010).

\bibitem{Ashhab2010} S. Ashhab and F. Nori, Phys. Rev. A {\bf 81}, 042311 (2010).

\bibitem{Braak2011} D. Braak, Phys. Rev. Lett. {\bf 107}, 100401 (2011).

\bibitem{Zhong} H. Zhong, Q. Xie, M. Batchelor, and C. Lee, J. Phys. A: Math. Theor. {\bf 46}, 415302 (2013).

\bibitem{Larson} J. Larson, J. Phys. B: At. Mol. Opt. Phys. {\bf 46}, 224016 (2013).

\bibitem{Tomka} M. Tomka, O. El Araby, M. Pletyukhov, and V. Gritsev, Phys. Rev. A {\bf 90}, 063839 (2014).

\bibitem{Li} Z.-M. Li and M. T. Batchelor, J. Phys. A: Math. Theor. {\bf 48}, 454005 (2015).

\bibitem{Forn2016} P. Forn-D\'iaz, G. Romero, C. J. P. M. Harmans, E. Solano, and J. E. Mooij, Sci. Rep. {\bf 6}, 26720 (2016).

\bibitem{Batchelor} M. T. Batchelor, Z.-M. Li, and H.-Q. Zhou, J. Phys. A: Math. Theor. {\bf 49}, 01LT01 (2016).

\bibitem{Yoshihara} F. Yoshihara, T. Fuse, S. Ashhab, K. Kakuyanagi, S. Saito, and K. Semba, Nature Phys. {\bf 13}, 44 (2017); Phys. Rev. A {\bf 95}, 053824 (2017).

\bibitem{Rossatto} D. Z. Rossatto, C. J. Villas-B\^oas, M. Sanz, and E. Solano, Phys. Rev. A {\bf 96}, 013849 (2017).

\bibitem{Wakayama} M. Wakayama, J. Phys. A: Math. Theor. {\bf 50}, 174001 (2017).

\bibitem{Kimoto} K. Kimoto, C. Reyes-Bustos, and M. Wakayama, arXiv:1712.04152.

\bibitem{Semple} J. Semple and M. Kollar, J. Phys. A: Math. Theor. {\bf 51} 044002 (2018).

\bibitem{Guan} K.-L. Guan, Z.-M. Li, C. Dunning, and M. T. Batchelor, J. Phys. A: Math. Theor. {\bf 51}, 315204 (2018).

\bibitem{Mao} B.-B. Mao, M. Liu, W. Wu, L. Li, Z.-J. Ying, and H.-G. Luo, Chinese Phys. B {\bf 27}, 054219 (2018).

\bibitem{Wang} G. Wang, R. Xiao, H. Z. Shen, C. Sun, and K. Xue, Sci. Rep. {\bf 9}, 4569 (2019).

\bibitem{Braak2019} D. Braak, Symmetry {\bf 11}, 1259 (2019).

\bibitem{Ashhab2008} S. Ashhab, A. O. Niskanen, K. Harrabi, Y. Nakamura, T. Picot, P. C. de Groot, C. J. P. M. Harmans, J. E. Mooij, and F. Nori, Phys. Rev. B {\bf 77}, 014510 (2008).

\end{thebibliography}
\end{document}